\documentclass[preprint,aps]{revtex4}

\usepackage{graphicx}

\begin{document}

\title{Strong Anisotropy of Dirac Cones in SrMnBi$_2$ and CaMnBi$_2$  Revealed by Angle-Resolved Photoemission Spectroscopy}

\author{Ya Feng$^{1}$, Zhijun Wang$^{1}$, Chaoyu Chen$^{1}$, Youguo Shi$^{1}$, Zhuojin Xie$^{1}$, Hemian Yi$^{1}$, Aiji Liang$^{1}$, Shaolong He$^{1}$, Junfeng He$^{1}$, Yingying Peng$^{1}$, Xu Liu$^{1}$, Yan Liu$^{1}$, Lin Zhao$^{1}$, Guodong Liu$^{1}$, Xiaoli Dong$^{1}$, Jun Zhang$^{1}$, Chuangtian Chen$^{3}$, Zuyan Xu$^{3}$, Xi Dai$^{1}$,, Zhong Fang$^{1}$, and X. J. Zhou$^{1,2,*}$}

\affiliation{
\\$^{1}$Beijing National Laboratory for Condensed Matter Physics, Institute of Physics, Chinese Academy of Sciences, Beijing 100190, China
\\$^{2}$Collaborative Innovation Center of Quantum Matter, Beijing, China
\\$^{3}$Technical Institute of Physics and Chemistry, Chinese Academy of Sciences, Beijing 100190, China.
\\$^{*}$Corresponding author: XJZhou@aphy.iphy.ac.cn.
}

\date{December 18, 2013}
%




\maketitle

{\bf
The Dirac materials, such as graphene and three-dimensional topological insulators, have attracted much attention because they exhibit novel quantum phenomena with their low energy electrons governed by the relativistic Dirac equations. One particular interest is to generate Dirac cone anisotropy so that the electrons can propagate differently from one direction to the other, creating an additional tunability for  new properties and applications. While various theoretical approaches have been proposed to make the isotropic Dirac cones of graphene into anisotropic ones, it has not yet been met with success.  There are also some theoretical predictions and/or experimental indications of anisotropic Dirac cone in novel topological insulators and AMnBi$_2$ (A=Sr and Ca) but more experimental investigations are needed.  Here we report systematic high resolution angle-resolved photoemission measurements that have provided direct evidence on the existence of strongly anisotropic Dirac cones in SrMnBi$_2$ and CaMnBi$_2$.  Distinct behaviors of the Dirac cones between SrMnBi$_2$ and CaMnBi$_2$ are also observed.  These results have provided important information on the strong anisotropy of the Dirac cones in AMnBi$_2$ system that can be governed by the spin-orbital coupling and the local environment surrounding the Bi square net.
}


The Dirac materials, so called because the behaviors of the low energy electrons in these materials can  be described by the relativistic Dirac equation, have recently attracted much attention \cite{OVafek}. Such an interest is first triggered mainly by the discovery of graphene with its characteristic Dirac cones near the zone corner\cite{NovoselovSci,RMPgraphene}. The interest is further  pushed to the full front by the discovery of the topological insulators with their characteristic Dirac cone in the topological surface state\cite{RevHasan,RevZSC}.  The Dirac materials have expanded to encompass many other materials including the high temperature cuprate superconductors with a nodal Dirac cone in the {\it d}-wave superconducting state\cite{OrensteinSci}, the parent compound of the iron-based compounds\cite{GDLiuFe,RichardFe} and silicene\cite{Vogt,LChen,BJFeng}.  Since the Dirac cone exhibits linear dispersion, massless and chiral low energy excitations\cite{OVafek}, the Dirac materials exhibit a number of exotic and novel quantum phenomena including quantum Hall effect\cite{QHEGraphene1,QHEGraphene2,Katsnelson,QED}.  In addition to searching for new Dirac materials\cite{XGWan,HgCrSe,Burkov,Na3Bi,Cd2As3}, a great effort has been devoted to engineer the Dirac cone.  One particular aspect is to generate Dirac cone anisotropy because anisotropic Dirac transport can be harnessed for making new electronic devices when electrons propagate differently from one direction to the other. To generate Dirac anisotropy, various approaches have been proposed but have not yet been materialized, including patterned periodic potential\cite{CHPark} or strain\cite{Grastrain}, heterostructures\cite{VPardo,Banerjee} and others\cite{Volovik,PDietl,AgTe,HgS,HgSe}. The AMnBi$_2$ (A=Sr or Ca) system has attracted special attention because it is expected that in this natural material, an anisotropic Dirac cone may be realized\cite{JPark,JKWang,KFWangSMB,KFWangCMB,JBHe,GLee}. In particular, band structure calculations\cite{JPark,GLee} and initial angle-resolved photoemission measurements on SrMnBi$_2$\cite{JPark} indicate that the anisotropy of the Dirac cone in AMnBi$_2$ can be governed by local arrangement of Sr (or Ca) surrounding the Bi square net.  This behavior, if fully proven and understood, will provide important information in understanding the origin of Dirac cone anisotropy and provide an ideal platform to tune the anisotropy of the Dirac cone.

Angle-resolved photoemission spectroscopy (ARPES) is a powerful tool which can directly reveal the presence of the Dirac cone and its anisotropy\cite{ADamascelli}. While there are some initial ARPES results for SrMnBi$_2$\cite{JPark}, so far no ARPES measurements have been reported for CaMnBi$_2$. In this paper, we present detailed high resolution angle-resolved photoemission results on both SrMnBi$_2$ and CaMnBi$_2$. Our results show that these two compounds are Dirac materials with strongly anisotropic Dirac cones.  We have found different behaviors of the Dirac cone between SrMnBi$_2$ and CaMnBi$_2$. In particular, in contrast to the band structure calculations before\cite{GLee}, we find that CaMnBi$_2$ is a Dirac material with isolated Dirac cones but with giant anisotropy, which is consistent with our band structure calculation results. These results have established a  clear case of Dirac materials with strongly anisotropic Dirac cones in AMnBi$_2$ (A=Sr and Ca) system.

High quality single crystals of SrMnBi$_2$ and CaMnBi$_2$ with a typical dimension of 3$\times$3 mm$^{2}$ were grown by self-flux method.  The ARPES measurements were carried out on our Lab photoemission system equipped with the Scienta R4000 electron energy analyzer and Helium discharge lamp which gives a photon energy of h$\upsilon$= 21.218 eV\cite{GDLiu}. The energy resolution was set at 20 meV and the angular resolution is $\sim$0.3 degree. The Fermi level is referenced by measuring on the Fermi edge of a clean polycrystalline gold that is electrically connected to the sample.  The crystals were cleaved {\it in situ} and measured in vacuum with a base pressure better than 5$\times$10$^{-11}$ Torr.

The electronic structure calculations were performed by using the full-potential augmented plane-wave and Perdew-Burke-Ernzerhofparametrization of the generalized gradient approximation(GGA-PBE) exchange-correlation function\cite{PBE} as implemented in the WIEN2k code. The spin-orbital interaction were included by using a second variational procedure. The muffin-tin radii R$_{MT}$ were set to 2.50 bohrs for all atoms. The plane-wave cut off K$_{max}$ was determined by R$_{min}$K$_{max}$=7.0, where R$_{min}$ is the minimal R$_{MT}$.A k-point mesh of 13$\times$13$\times$5 is used for both materials. Our calculated results are consistent with the previous report\cite{GLee}.

Figure 1 shows the constant-energy contours of SrMnBi$_2$ and CaMnBi$_2$ at different binding energies.  The corresponding electronic structures of SrMnBi$_2$ and CaMnBi$_2$ are presented in Fig. 2 and Fig. 3, respectively.  For SrMnBi$_2$, there are two main features observed in the measured Fermi surface (Fig. 1c1). One is the weak square-shaped hole-like Fermi surface around $\Gamma$. The other is the four separated strong intensity segments in the first Brillouin zone that are confined along the $\Gamma$-M directions.  Each segment grows in area with the increasing binding energy and gradually becomes a crescent moon-like shape (Fig. 1c2-c5). The Fermi surface of CaMnBi$_2$ (Fig. 1d1) exhibits similar features in that it also shows a hole-like square-shaped Fermi surface around $\Gamma$  with a relatively strong intensity.  But it shows a large diamond-like Fermi surface connecting four equivalent X points in the first Brillouin zone, different from the four isolated segments in SrMnBi$_2$.  With the increase of the binding energy, the portion of the large continuous contour near the $\Gamma$-M region also increases in area and forms  four crescent moon-like pockets (Fig. 1d2-d5).

Figure 2 shows a detailed evolution of electronic structure with momentum in SrMnBi$_2$.   For each momentum cut, there are two sets of bands. One is a slightly broad inner band, denoted as IB in Fig. 2b for cut A, that gives rise to the square-shaped Fermi surface around $\Gamma$ (Fig. 1c1). The other is an outer band, denoted as DB in Fig. 2b for cut A,  that leads to the four strong-intensity segments in Fig. 1c1.  The outer DB band consists of two sharp linear bands, observed for the momentum cuts A-F in Fig. 2a,  as marked by the red dashed lines in Fig. 2b. They can be seen more clearly in Fig. 2c which give an expanded view of the left-side Dirac band in the second derivative images of the original data (Fig. 2b). The outer side of the DB linear band extends to high binding energy while the inner side becomes invisible when it merges with the inner IB bands.  It is clear that the crossing point of the two DB linear bands is highest along the $\Gamma$-M  direction (cut A) and lies above the Fermi level E$_F$ for the cuts A, B and C (Fig. 2b and 2c).  When the momentum cuts move away from the $\Gamma$-M direction, the energy position of the crossing point goes downwards and sinks to below E$_F$ for the cuts D, E and F.  We note that our results are different from the previous report where the Dirac point is below the Fermi level\cite{JPark}. This difference may come from doping difference during the sample preparation. In order to quantitatively determine the location of the DB band crossing point in the momentum space and its corresponding energy, the two sides of the DB band are represented by two straight lines and the intersection points are taken as the band crossing position (dashed lines in Fig. 2b).

Figure 3 shows eletronic structure evolution with momentum in CaMnBi$_2$. It also consists of the inner IB bands and outer DB Dirac bands, as labeled in Fig. 3b for the cut C. However, the Dirac bands in CaMnBi$_2$ exhibit quite different behaviors from that in SrMnBi$_2$. As seen from Fig. 3b, for the momentum cuts A and B near the $\Gamma$-M direction, the DB band crossing point lies slightly above the Fermi level, which can be more clearly seen in the expanded view of bands in Fig. 3c.  When the momentum cut moves away from the $\Gamma$-M direction, the DB band crossing point also shows a similar drop as in SrMnBi$_2$. But another band shows up near the Fermi level and becomes obvious for the cut G in Fig. 3c.  As it is shown below from the band structure calculations (Fig. 4), because of the strong spin-orbital coupling, the Dirac band is gapped forming the upper Dirac and lower Dirac branches. In SrMnBi$_2$, for all the momentum cuts (A-F), we observe only the lower Dirac branch (Fig. 2b and 2c). But in CaMnBi$_2$, in addition to the lower Dirac branch for all the cuts, we also observe the upper Dirac branch for the cuts F and G (Fig. 3b and 3c). There is a gap opening between the upper and lower Dirac branches signalled as suppressed spectral weight between them. These observations explain why we can see a large diamond-like Fermi surface in CaMnBi$_2$ because both the upper and lower Dirac band branches cross the Fermi level.  It is also consistent with the spectral weight variation along the ``underlying Fermi surface"  in CaMnBi$_2$ (Fig. 3a). One can see strong spectral weight near the $\Gamma$-M region (cuts A, B, C and D) (lower Dirac band branch crosses or touches the Fermi level), suppressed spectral weight near the cut E (a gap forms between the upper and lower Dirac band branches at the Fermi level), and  strong spectral weight again near X region (cuts F and G) (upper Dirac branch crosses the Fermi level).

The strong spin-orbital coupling in SrMnBi$_2$ and CaMnBi$_2$ produces not only a gap opening of the Dirac band, but also a band asymmetry of the Dirac cone, as seen in Fig. 4. The signature of gap opening is already observed in CaMnBi$_2$ (Fig. 3b and 3c).  Fig. 4a shows the Dirac band of SrMnBi$_2$ measured along $\Gamma$-M direction (cut G in Fig. 4a). It is clear that the Dirac band exhibits a clear asymmetry with the left-side band being much steeper than the right-side band (closer to $\Gamma$ point); the corresponding Fermi velocities of the left- and right-side bands are 10.9 eV$\cdot$$\AA$ and 2.4 eV$\cdot$$\AA$, respectively. The band structure calculations indicate clearly that such a Dirac band asymmetry is induced by the strong spin-orbital coupling in SrMnBi$_2$.  As seen in Fig. 4b, without considering the spin-orbital coupling, the calculated Dirac band (dashed pink lines) is symmetrical without gap opening. Upon introducing  a strong spin-orbital coupling in the band structure calculations, the Dirac band (thick blue lines) splits from the Dirac point to form a gap between the upper and lower Dirac band branches, accompanied by the formation of band asymmetry. Our calculations are similar to that reported before\cite{GLee}. Similar behavior is also observed in CaMnBi$_2$ where the Fermi velocity of the left and right-side bands of the lower Dirac branch (Fig. 4c, measured along $\Gamma$-M for the cut H in Fig. 3a)) are  10.6 eV$\cdot$$\AA$  and 2.1 eV$\cdot$$\AA$, respectively. The calculated result for CaMnBi$_2$ (Fig. 4d) is similar to that in SrMnBi$_2$ (Fig. 4c) for the Dirac band along the $\Gamma$-M direction.


Figure 5 summarizes the momentum locus of the crossing points obtained from different momentum cuts in SrMnBi$_2$ (Fig. 2) and CaMnBi$_2$ (Fig. 3),  their corresponding energy position along the locus and the Dirac band dispersion along the $\Gamma$-M direction. For SrMnBi$_2$, since the upper Dirac branch is above the Fermi level for all the momentum cuts, we can determine the momentum crossing point (Fig. 5a) and its corresponding energy (Fig. 5c) only for the lower Dirac branch.  It is clear that, along the $\Gamma$-M direction, it shows a steep Dirac dispersion (thick black line in Fig. 5c) with a Fermi velocity for the left-side band as 10.9 eV$\cdot$$\AA$.  Along the momentum locus near the diagonal region (perpendicular to the $\Gamma$-M direction), the dispersion becomes rather flat, with a Fermi velocity of $\sim$0.5 eV$\cdot$$\AA$ near the Dirac point. The dramatic difference between the Fermi velocity along $\Gamma$-M and perpendicular to $\Gamma$-M directions points to a strong anisotropy of the Dirac cone in SrMnBi$_2$.   For CaMnBi$_2$, while the crossing points near $\Gamma$-M direction are above the Fermi level, away from the $\Gamma$-M direction and close to the X point, it is possible to observe both the lower branch and the upper branch, and the gap opening due to the spin-orbital coupling. The crossing points of the lower Dirac branch (cuts A, B, C and D in Fig. 3) and the upper Dirac branch (cuts E, F and G) are plotted in Fig. 5b.  The corresponding energy position of the crossing points for both the lower and upper Dirac branches are shown in Fig. 5d. Along the $\Gamma$-M direction, the Dirac band is also very steep (thick black lines in Fig. 5d), with a Fermi velocity for the left-side band being  10.6 eV$\cdot$$\AA$. But along the momentum locus (perpendicular to the $\Gamma$-M direction), the dispersion becomes rather flat over a relatively large momentum space with a Fermi velocity of $\sim$0.1 eV$\cdot$$\AA$ near the diagonal region. This proves that,  in CaMnBi$_2$, the anisotropy of the Dirac cone is even stronger than that in SrMnBi$_2$. We note that, because of the arrangement of Ca above and below the Bi square net (Fig. 1b),  an almost continuous band crossing line with all the Dirac crossing points at the same energy was expected for CaMnBi$_2$\cite{GLee}. However, our measurements clearly show that this is not the case because the drop of the crossing point is clear from cut A to cut C (Fig. 3c). This indicates that CaMnBi$_2$ still has isolated Dirac cones, which is consistent with our band structure calculations below (Fig. 6).

While SrMnBi$_2$ and CaMnBi$_2$ exhibit some qualitative similarities on the anisotropic Dirac cone structure as shown above, there are a number of obviously different behaviors between them. First, the measured Fermi surface is different; while SrMnBi$_2$ shows four discrete strong intensity segments in the first Brillouin zone (Fig. 1c1), CaMnBi$_2$ exhibits a large diamond-like Fermi surface (Fig. 1d1).  Second, in SrMnBi$_2$, only the lower Dirac branch is observed (Fig. 2b and 2c), while in CaMnBi$_2$, both the lower and upper Dirac branches can be seen (Fig. 3b and 3c). Third, the band dispersion along the momentum locus near the $\Gamma$-M diagonal region is much flat in CaMnBi$_2$ (Fig. 5d) than that in SrMnBi$_2$ (Fig. 5c). These behaviors can be well understood from the band structure calculations including the spin-orbital coupling (Fig. 6).  As seen from Fig. 6, the effect of the spin-orbital coupling on the Dirac cone is quite different between SrMnBi$_2$ and CaMnBi$_2$. While the gap opening along $\Gamma$-M direction is comparable, it is much larger in SrMnBi$_2$ (Fig. 6a) than that in CaMnBi$_2$ (Fig. 6d) near X region (highlighted by blue circle). This explains why, in SrMnBi$_2$, we can only observe the lower Dirac branch because the gap is large and the upper Dirac branch stays above the Fermi level. In contrast, in CaMnBi$_2$, since the gap opening is small, we can observe that the upper Dirac branch changes from above the Fermi level near the $\Gamma$-M direction to below the Fermi level near the X region (Fig. 3 and Fig. 6d). Such a different band evolution leads to different Fermi surface topology seen in SrMnBi$_2$ and CaMnBi$_2$.  Fig. 6b and Fig. 6e show constant energy contours for the Dirac bands; their corresponding three-dimensional images are shown in Fig. 6c and Fig. 6f, respectively. These calculations show a good agreement with our measurements (Fig. 1c and 1d), in particular, the crescent-moon-like Fermi pocket along the diagonal region. From these constant energy contours, we can also obtain the calculated momentum locus for SrMnBi$_2$ (solid yellow line in Fig. 5c) and CaMnBi$_2$ (solid yellow line in Fig. 5d). They show a good agreement with the measured results.

The distinct behaviors observed between SrMnBi$_2$ and CaMnBi$_2$ have provided important information that an anisotropic Dirac cone can be generated and manipulated in the AMnBi$_2$ (A: alkaline earth like Sr and Ca) system.  Band structure calculations have shown that the Dirac cone in AMnBi$_2$ (A=Sr or Ca) mainly originates from the A-Bi-A blocks in the crystal structure where the Bi square net is sandwiched between the upper and lower A layers (Figs. 1a and 1b)\cite{JPark,JKWang,GLee}.  The Dirac bands come mainly from the Bi p$_{x,y}$ orbitals in the Bi square net which are weakly hybridized with Sr or Ca d orbitals. The single-layer Bi square net can produce a continuous Dirac line\cite{GLee}. The effect of the introduction of A ions above and below the Bi square net (Fig. 1a and 1b) is two fold. First, the particular arrangement of A ions has different effect on lifting the degeneracy of the initial Dirac line from the Bi square net. In this regard, it has been shown that the coincident arrangement of Sr above and below Bi square net in SrMnBi$_2$ (Fig. 1a) has a larger effect than the staggered arrangement of Ca in CaMnBi$_2$ (Fig. 1b)\cite{GLee}. These are consistent with our observation that the dispersion along the momentum locus is much flatter in CaMnBi$_2$ (Fig. 5d) than that in SrMnBi$_2$ (Fig. 5c).  Second, as shown from the band structure calculations, different A ions gives rise to different spin-orbital coupling effect. Therefore, the selection of A ions provides a good handle in tuning the Dirac cone structure in the AMnBi$_2$ system.

In summary, through our detailed high resolution ARPES measurements, we have shown that SrMnBi$_2$ and CaMnBi$_2$ are Dirac materials with highly anisotropic Dirac cones.  We have revealed the difference of the Dirac cone structure between SrMnBi$_2$ and CaMnBi$_2$ that originates from the combined effect of  the spin-orbital coupling  and the particular arrangement of A ions above and below the Bi Square net.   The Bi square net in AMnBi$_2$ (A=Sr and Ca) provides an ideal platform from which, by tuning the surrounding environment and spin-orbital coupling, one can engineer the anisotropy of the Dirac cones\cite{GLee}.  The observation of anisotropic Dirac cones will also help in understanding the unusual properties of the (Sr,Ca)MnBi$_2$ system\cite{JPark,JKWang,KFWangSMB,KFWangAPL,KFWangCMB,JBHe}.\\

\vspace{3mm}

\textbf{Acknowledgments}

 This work is supported by the National Natural Science Foundation of China (91021006, 10974239, 11174346 and 11274367) and the Ministry of Science and Technology of China (2011CB921703, 2013CB921700 and 2013CB921904).

\vspace{3mm}

\textbf{Author contributions}
Y.F. and X.J.Z. conceived and designed the research; Y.F. performed measurements with C.Y.C., Z.J.X., H.M.Y., A.J.L., S.L.H.; Z.J.W, X.D. and Z. F. carried out the band structure calculations. Y.G.S. prepared the samples;  Y.F., C.Y.C., Z.J.X., H.M.Y., A.J.L., S.L.H., J.F.H., Y.Y.P., X.L., Y.L., L.Z., G.D.L., X.L.D., J.Z., C.T.C., Z.Y.X. and X.J.Z. contributed new analytic tools; Y.F., C.Y.C. and X.J.Z. analyzed data and wrote the paper. All authors participated in the discussion and comment on the paperer.

\vspace{3mm}
\textbf{Additional information}

\vspace{3mm}
\textbf{Competing financial interests}:
 The authors declare no competing financial interests.

\newpage

\begin{figure*}[t]
\begin{center}
\includegraphics[width=0.99\columnwidth,angle=0]{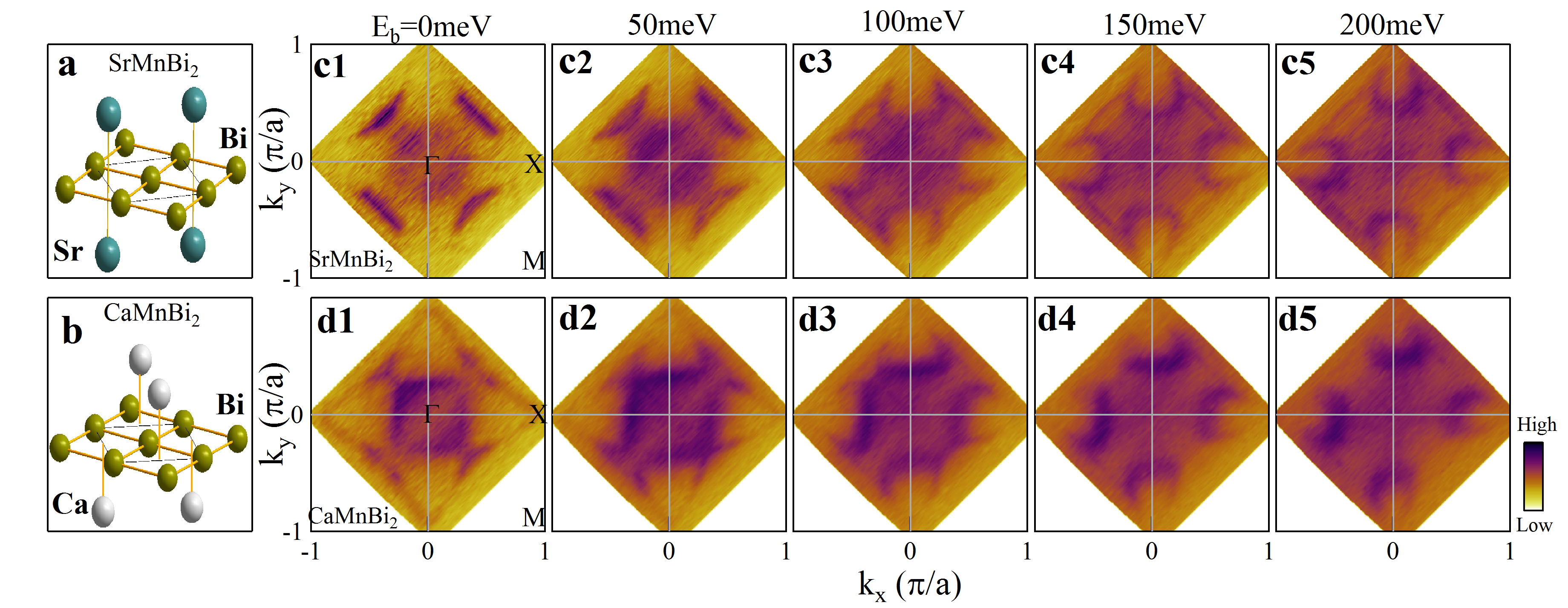}
\end{center}
\begin{center}
\caption{Constant energy contours for SrMnBi$_2$ and CaMnBi$_2$ measured at 30 K.  (a) and (b) show the Bi square net sandwiched between two Sr layers in SrMnBi$_2$ and two Ca layers in CaMnBi$_2$ in their crystal strucrures\cite{GLee}.  (c1-c5). Constant energy contours of SrMnBi$_2$ obtained by integrating the photoemission spectral weight over a 10 meV energy window with respect to the binding energy of 0 (c1), 50 meV (c2), 100 meV (c3), 150 meV (c4) and 200 meV (c5).  The crystal structure of SrMnBi$_2$ is tetragonal with a space group of I4/mmm and  a lattice constant of a=4.58 $\AA$\cite{GLee}.  (d1-d5). Constant energy contours of CaMnBi$_2$ obtained by integrating the photoemission spectral weight over a 10 meV energy window with respect to the binding energy of 0 (d1), 50 meV (d2), 100 meV (d3), 150 meV (d4) and 200 meV (d5). The crystal structure of CaMnBi$_2$ is tetragonal with a space group of P4/nmm and a lattice constant of a=4.50 $\AA$\cite{GLee}. All the contours are obtained by symmetrizing the original data with respect to the ($\pi$,-$\pi$)-(-$\pi$,$\pi$) line.
}
\end{center}
\end{figure*}

\begin{figure*}[t]
\begin{center}
\includegraphics[width=0.99\columnwidth,angle=0]{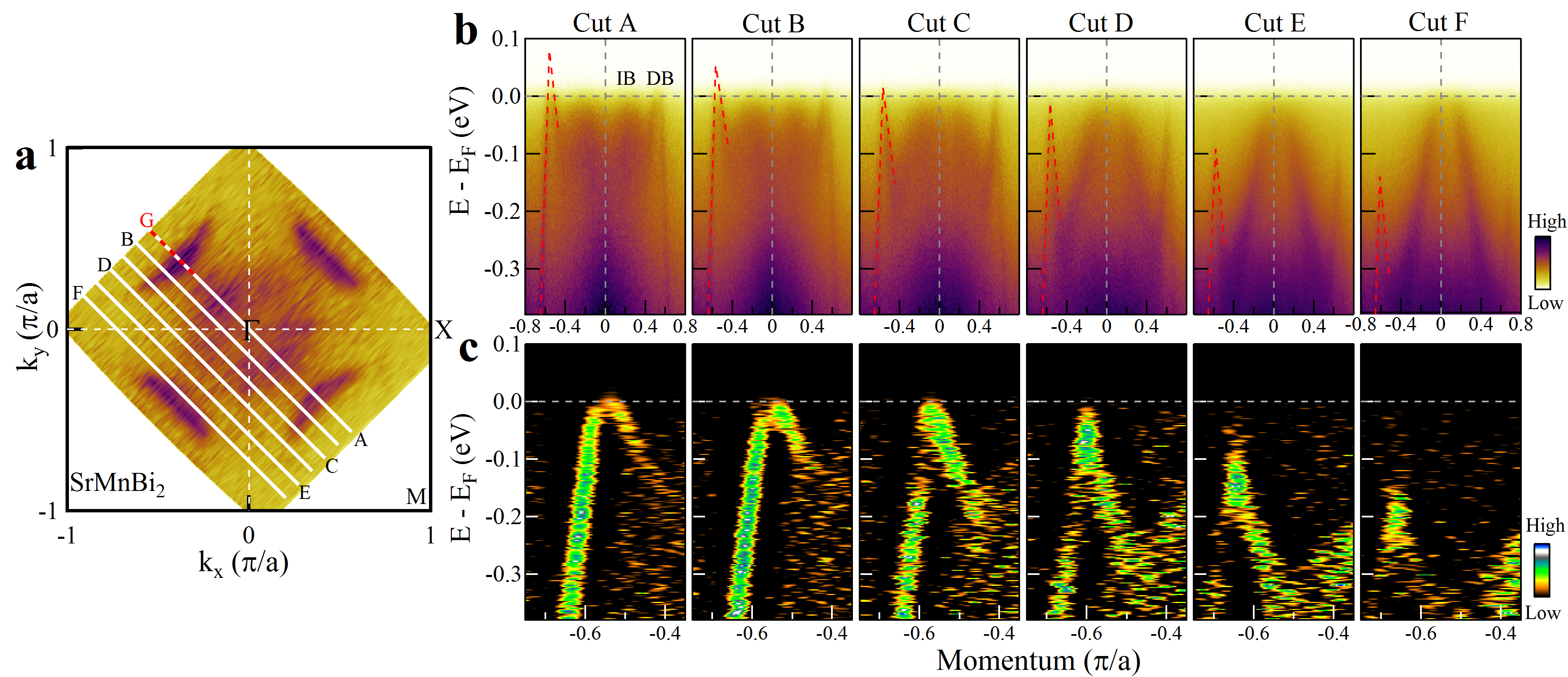}
\end{center}
\begin{center}
\caption{Momentum dependence of the electronic structure in SrMnBi$_2$. (a). Fermi surface of SrMnBi$_2$ measured at 30 K with the momentum cuts labeled from A to F. (b). Electronic structure along various momentum cuts labeled in (a) measured at 90 K. A relatively high temperature of 90 K is used here in order to see part of the bands above the Fermi level.  Dashed red lines are drawn, as a guide to eyes, on top of one Dirac band on the left side of the images. (c) Expanded view of the Dirac bands on the left side of (b). To highlight the electronic structure, these images are the second derivative of the original images in (b) with respect to the momentum.
}
\end{center}
\end{figure*}

\begin{figure*}[tbp]
\begin{center}
\includegraphics[width=0.99\columnwidth,angle=0]{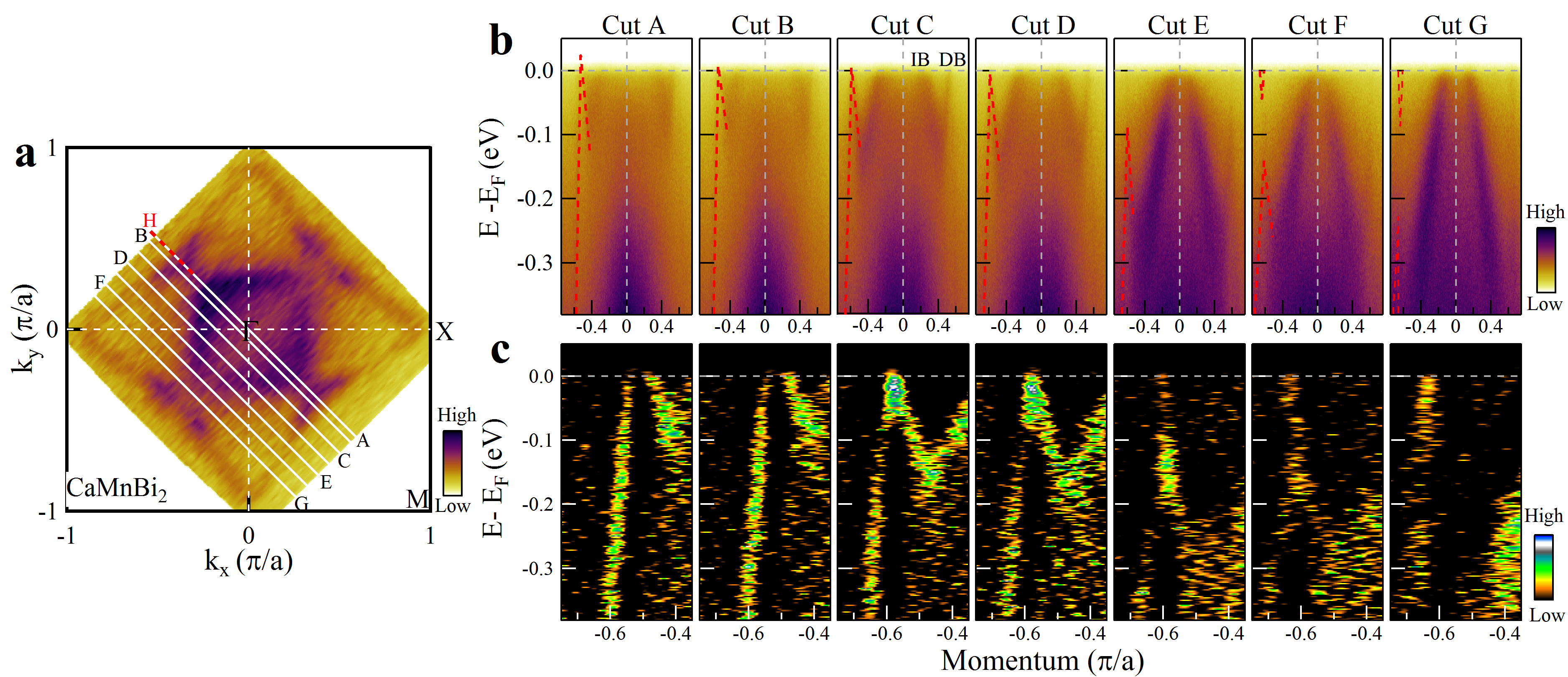}
\end{center}
\begin{center}
\caption{Momentum dependence of the electronic structure in CaMnBi$_2$. (a). Fermi surface of CaMnBi$_2$ measured at 30 K with the momentum cuts labeled. (b). Electronic structure along various momentum cuts labeled in (a) measured at 30 K. Dashed red lines are drawn, as a guide to eyes, on top of one Dirac band on the left side of the images. (c) Expanded view of the Dirac bands on the left side of (b). To highlight the electronic structure, these images are the second derivative of the original images in (b) with respect to the momentum.
}
\end{center}
\end{figure*}

\begin{figure*}[tbp]
\begin{center}
\includegraphics[width=0.99\columnwidth,angle=0]{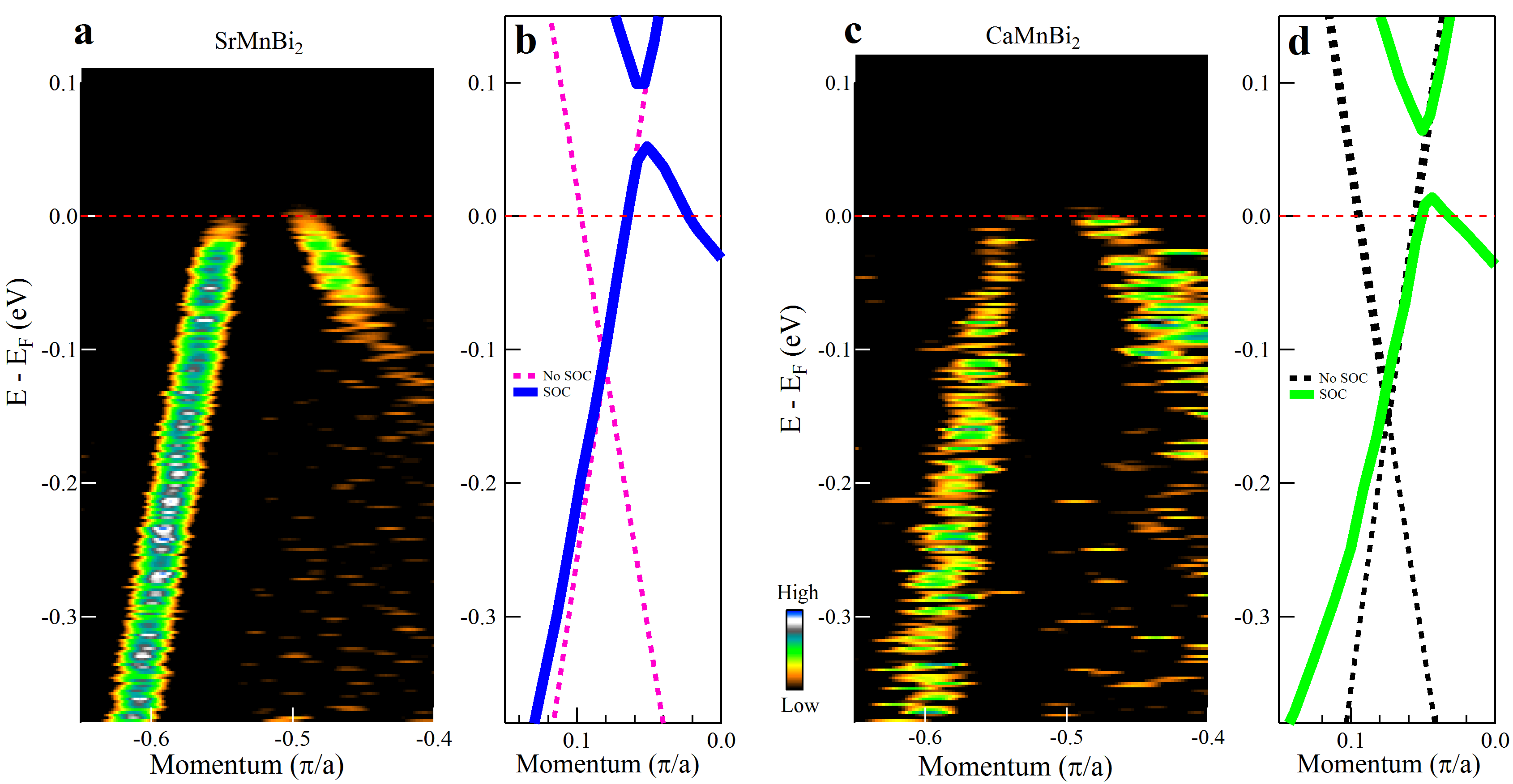}
\end{center}
\begin{center}
\caption{Effect of the spin-orbital coupling on the Dirac band in SrMnBi$_2$ and CaMnBi$_2$. (a). Asymmetric Dirac band measured on SrMnBi$_2$ along the $\Gamma$-M direction (cut G in Fig. 2a, dashed red line). It is the second derivative image with respect to the momentum. (b). Calculated band structure across the Dirac cone along the $\Gamma$-M direction in SrMnBi$_2$ without considering spin-orbital coupling (dashed pink lines) and with spin-orbital coupling (thick blue lines).  (c). Asymmetric Dirac band measured on CaMnBi$_2$ along the $\Gamma$-M direction (cut H in Fig. 3a, dashed red line). It is the second derivative image with respect to the momentum. (d). Calculated band structure across the Dirac cone along the $\Gamma$-M direction in CaMnBi$_2$ without considering spin-orbital coupling (dashed black lines) and with spin-orbital coupling (thick green lines).
}
\end{center}
\end{figure*}

\begin{figure*}[tbp]
\begin{center}
\includegraphics[width=0.99\columnwidth,angle=0]{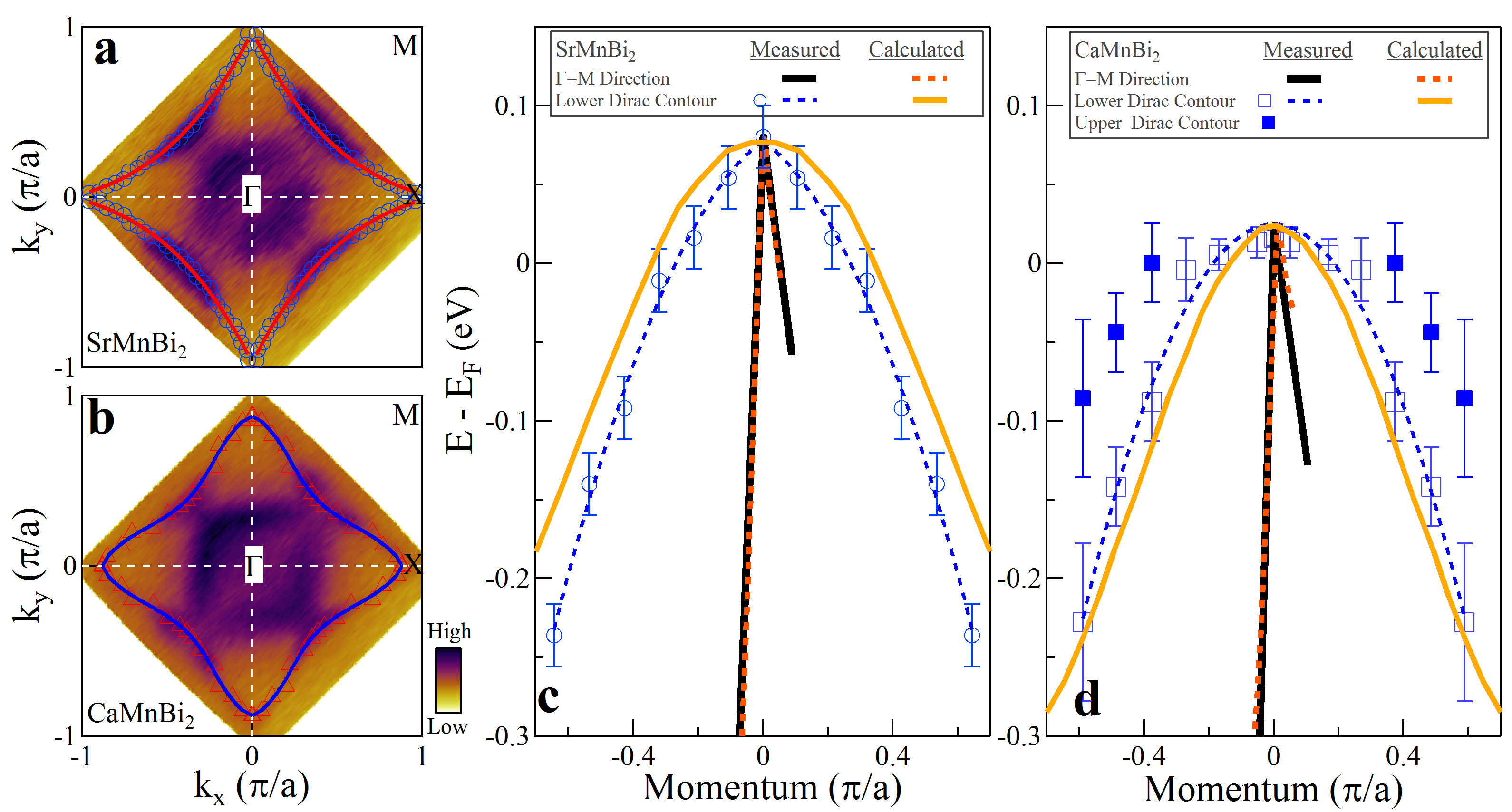}
\end{center}
\begin{center}
\caption{Dirac cone anisotropy in SrMnBi$_2$ and CaMnBi$_2$. (a) and (b) show locus of the crossing points in the momentum space in SrMnBi$_2$ (blue empty circles) and CaMnBi$_2$ (red empty triangles), respectively. (c). Measured lower Dirac band dispersion along the $\Gamma$-M direction (thick black line) and momentum dependence of the crossing point energy (empty blue circles, the dashed blue line is a guide to the eyes) along the underlying locus shown in (a)(red thick line) for SrMnBi$_2$. The corresponding calculated Dirac band along the $\Gamma$-M direction (red dashed line) and the cross point energy along the underlying locus (solid yellow line) are also plotted for comparison.   (d). Measured lower Dirac band dispersion along the $\Gamma$-M direction (thick black line) and momentum dependence of the lower-band-top energy (empty blue squares, the dashed blue line ia a guide to the eyes) and the upper-band-bottom (solid blue squares) along the underlying locus shown in (b)(blue thick line) for CaMnBi$_2$.   The corresponding calculated Dirac band along the $\Gamma$-M direction (red dashed line) and the cross point energy along the underlying locus (solid yellow line) are also plotted for comparison.
}
\end{center}
\end{figure*}

\begin{figure*}[tbp]
\begin{center}
\includegraphics[width=0.99\columnwidth,angle=0]{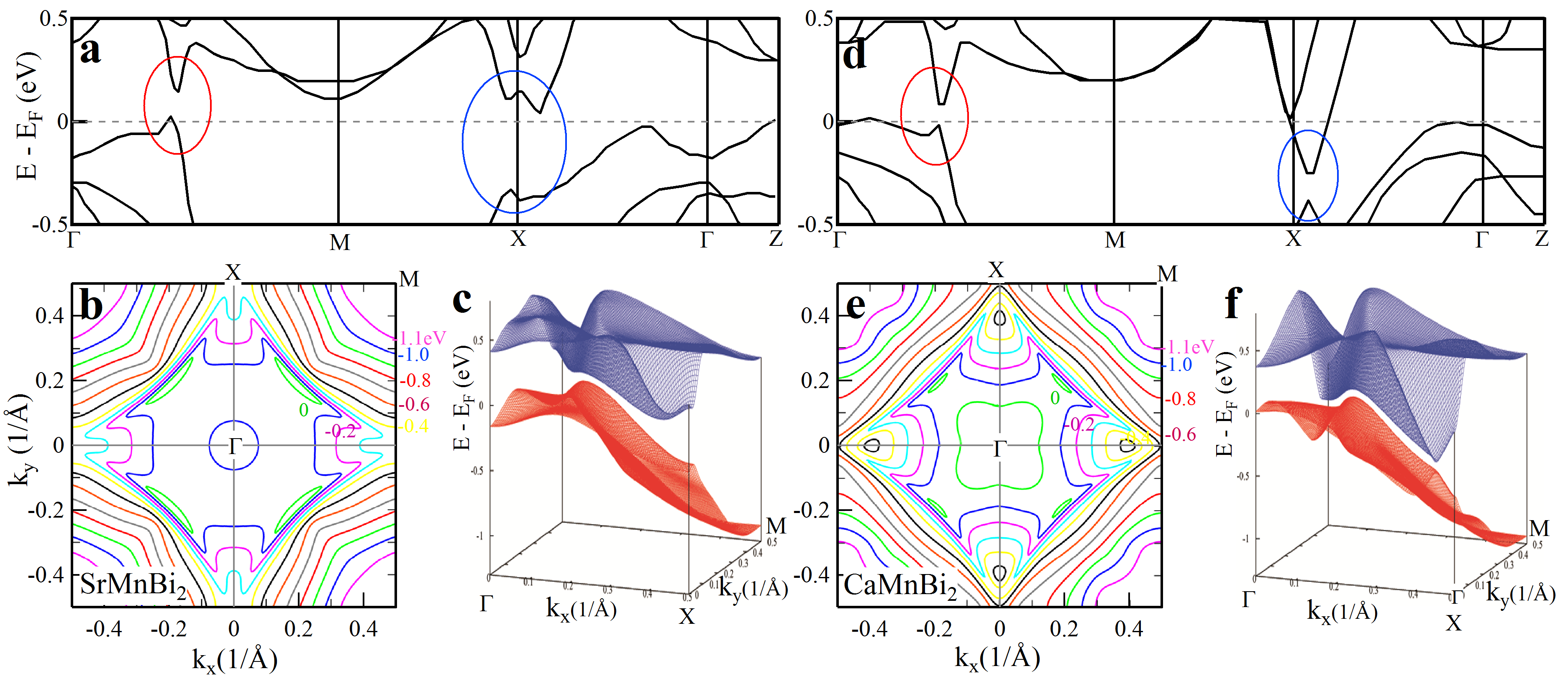}
\end{center}
\begin{center}
\caption{The Dirac cone structure in SrMnBi$_2$ and CaMnBi$_2$ from the band structure calculations. (a). Calculated Band structure of SrMnBi$_2$ along high symmetry directions considering the spin-orbital coupling and the checkboard-type antiferromagnetic order. Red circle and blue circle highlight the Dirac band features along  $\Gamma$-M and near X, respectively.  (b). Constant energy contours of SrMnBi$_2$ band structure at different energies. (c). Schematic band structure for SrMnBi$_2$ of the upper Dirac band (blue) and lower Dirac band (red) in one Brillouin zone quadrant.  (d). Calculated Band structure of CaMnBi$_2$ along high symmetry directions considering the spin-orbital coupling and the checkboard-type antiferromagnetic order. Red circle and blue circle highlight the Dirac band features along  $\Gamma$-M and near X, respectively.  (b). Constant energy contours of CaMnBi$_2$ band structure at different energies. (c). Schematic band structure for CaMnBi$_2$ of the upper Dirac band (blue) and lower Dirac band (red) in one Brillouin zone quadrant. }
\end{center}
\end{figure*}

\end{document}